# On the Role of Interlayer Electrons on the Frictional Behavior of Two-Dimensional Electrides

*Jingcheng Qi,[1] Giuliana Materzanini,[2] Gian-Marco Rignanese,[1,2,3] Maria Clelia Righi,[4] Junjie Wang,[1,*]*

[1]State Key Laboratory of Solidification Processing, Northwestern Polytechnical University, Xi'an 710072, People's Republic of China

[2]Institute of Condensed Matter and Nanosciciences (IMCN), Université Catholique de Louvain, Louvain-la-Neuve 1348, Belgium

[3]WEL Research Institute, Avenue Pasteur 6, 1300 Wavre, Belgium

[4]Department of Physics and Astronomy, University of Bologna, Bologna 40127, Italy

*wang.junjie@nwpu.edu.cn

**Abstract**

Friction accounts for up to 30% of global energy consumption, underscoring the urgent need for superlubricity in advanced materials. Two-dimensional (2D) electrides are layered materials with cationic layers separated by 2D confined electrons that act as anions. This study reveals the unique frictional properties of these compounds and the underlying mechanisms. We establish that interlayer friction correlates with the cationic charges and sliding-induced charge redistribution. Remarkably, the 2D electride $Ba_2N$ stands out for its lower interlayer friction than graphene, despite its stronger interlayer adhesion, defying conventional tribological understanding. This anomalous behavior arises from electron redistribution as the dominant energy dissipation pathway. Combining *ab initio* calculations and deep potential molecular dynamics (DPMD) simulations, we show that incommensurate twisted interfaces ($2° < \theta < 58°$) in $Ba_2N$ achieve structural superlubricity by suppressing out-of-plane buckling and energy corrugation. Notably, a critical normal load of 2.3 GPa enables barrier-free sliding in commensurate $Ba_2N$ ($\theta = 0°$), with an ultralow shear-to-load ratio of 0.001, suggesting the potential for superlubricity. Moreover, electron doping effectively reduces interlayer friction by controllably modulating stacking energies in 2D electrides. These findings establish 2D electrides as a transformative platform for energy-efficient tribology, enabling scalable superlubricity through twist engineering, load adaptation, or electrostatic gating. Our work advances the fundamental understanding of electron-mediated friction, with $Ba_2N$ serving a model system for cost-effective, high-performance material design.

## Introduction

Friction, a widespread physical phenomenon, frequently results in undesirable consequences such as wear failures and energy loss across industries, with research suggesting that up to 30% of energy consumption can be attributed to it[1, 2]. Consequently, significant efforts have been directed towards minimizing friction and achieving "superlubricity", characterized by extremely low friction values. In 2004, Frenken et al. experimentally demonstrated superlubricity in graphite[3]. They observed that frictional forces between a graphite sheet and a highly oriented pyrolytic graphite (HOPG) crystal surface are significant only when the graphite sheet has a commensurate orientation relative to the HOPG substrate. In contrast, friction is significantly reduced for incommensurate orientations, a phenomenon referred to as "structural lubricity"[4]. Superlubricity was also achieved in commensurate contacts by applying specific external loads and strains[5, 6, 7]. Researchers continue to explore new materials and approaches that offer increased stability and reduced cost for achieving superlubricity[8, 9, 10, 11, 12, 13, 14, 15].

In atomic force microscopy (AFM) measurements, friction primarily originates from the out-of-plane motion of two-dimensional (2D) materials, which stems from the ultralow interlayer adhesion in van der Waals interface[16, 17]. Hence, enhancing substrate adhesion is key to reducing friction by suppressing out-of-plane wrinkling[18, 19]. It is essential to find materials that simultaneously possess high interlayer adhesion and low shear strength.

2D electrides, characterized by cationic layers hosting interlayer electrons (see **Fig. 1a**), have garnered interest for their potential applications in various fields.[20, 21, 22] The alternating cationic layers and interlayer electrons allow for a high interlayer binding energy than graphene (e.g., 1.11 J/m$^2$ for Ca$_2$N and 0.31 J/m$^2$ for graphite)[23], but a considerable separation (e.g., ~3.85 Å in Ca$_2$N[21]), comparable to typical van der Waals (vdW) distances.[24] Moreover, the high-density interlayer electrons exhibit remarkable activity, making it convenient to control their distribution. Previous studies on both Ca$_2$N bilayers and bulk materials have demonstrated that their electronic structure remains insensitive to stacking arrangements.[25] This phenomenon was simply attributed to the effective screening of cationic layers by interlayer electrons, which minimizes interlayer charge variations during sliding. However, it remains unclear whether a superlubricity state can be induced in 2D electrides. While previous studies have focused on charge screening effects, the fundamental relationship between electron redistribution, energy barriers, and friction dynamics remains unexplored. This knowledge gap motivates our systematic investigation of the friction behavior in 2D electrides.

In this study, we conduct *ab initio* calculations to thoroughly investigate the frictional properties of 2D electrides by analyzing their sliding potential energy surfaces (PES). We identify the crucial role of structural parameters and charge redistribution in determining the interlayer friction of layered electrides (see **Fig. 1b**). Surprisingly, the Ba$_2$N bilayer exhibits significantly lower intrinsic friction than graphite. Our deep potential molecular dynamics (DPMD) simulations demonstrate that suitable twist angles (2°<$\theta$<58°) can effectively suppresses the out-of-plane deformation in Ba$_2$N bilayer, enabling structural superlubricity (see **Fig. 1c**). Moreover, the commensurate ($\theta$=0°) Ba$_2$N exhibits ultralow interlayer friction under an extra normal load, with the ratio between the shear strength and normal load reaching 0.001, highlighting its potential for superlubricity (see **Fig. 1d**). Additionally, electron doping effectively lowers the interlayer friction in 2D electrides by modulating the energy differences between various stacking configurations (see **Fig. 1e**).

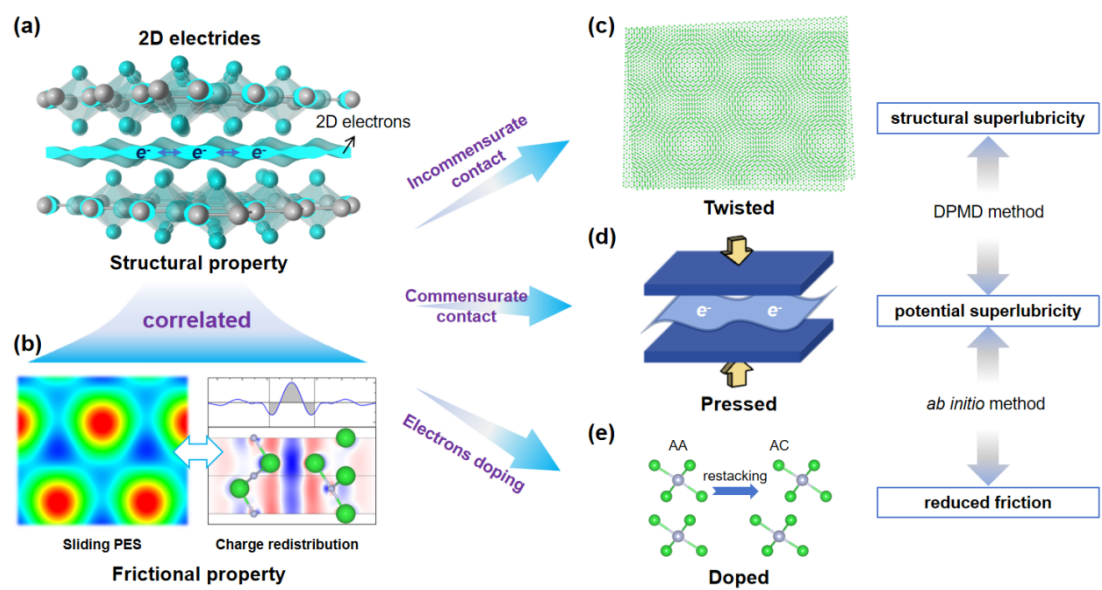

**Fig. 1**. Investigation of intrinsic frictional properties and friction-reduction strategies in 2D electrides. (a) Interlayer structure of 2D electrides characterized by localized 2D electrons distributed between two cationic layers. (b) Sliding PES and interlayer charge redistribution are employed to describe the influence of sliding processes on interlayer interactions. (c) Twisting-induced incommensurate contact of superlattice leads to structural superlubricity. (d) Out-of-plane pressure causes collapse of the sliding energy barrier, resulting in ultralow friction. (e) Electron-doping-induced interlayer restacking reduces the interlayer friction.

**Results**

**Structural and frictional properties of 2D electrides.** We investigate a series of $M_2X$ compounds with the anti-$CdCl_2$ structure (space group *R-3m,* #166), using alkaline-earth cations (M = Ca, Sr, and Ba) and nonmetal V-A elements (X = N, P, and As) as anions. Electrides like $Ca_2N$[21] and $Ba_2N$[26] have been confirmed experimentally, while others such as $Sr_2N$, $Sr_2P$, $Ba_2P$, and $Ba_2As$ have been predicted as stable 2D electrides through *ab initio* calculations.[27, 28] Although $Ca_2P$, $Ca_2As$, and $Sr_2As$ are thermodynamically metastable,[27] they are included for comparison. We fully optimize the conventional bulk unit cells of these compounds (see **Fig. 2a**), obtaining relaxed geometries and binding energies in agreement with previous reports (see Supplementary **Table 1**). The crystal structures feature three distinct layers, each composed of two atomic planes of M cations and one atomic plane of X anions. These layers are denoted as A, B, and C. The electron localization function (ELF) is calculated to visualize the electron distribution in the layers, with that of $Ca_2N$ shown in **Fig. 2a**. The interlayer electron sheets, spaced ~6.33 Å apart, show negligible interaction, thus a bilayer model—consisting of a single electron sheet flanked by two cationic layers and vacuum—adequately represents their bonding and frictional properties. Supplementary **Fig. 1** illustrates bilayer configurations with AB and AC stackings, related to one another by sliding the layers in the *x-y* plane, as detailed in the Supporting Information. For a general $M_2X$ compound, the bilayer configurations are constructed as follows: the atomic ($x$, $y$, $z$) coordinates of the lower layer are fixed, while in the upper layer, the $z$-coordinates are held at the values corresponding to the minimum-energy stacking configuration. The ($x$, $y$) coordinates of the upper layer are then rigidly displaced by a vector ($x^*$, $y^*$) relative to the minimum-energy configuration.

We calculate the sliding PES per area, $\Delta E_{(x^*, y^*)}$, defined as:

$$\Delta E_{(x^*, y^*)} = (E_{(x^*, y^*)} - E_{(0, 0)})/A \qquad (1)$$

where $E_{(x^*, y^*)}$ and $E_{(0, 0)}$ are the total energies of the ($x^*$, $y^*$)-displaced configuration and the minimum-energy stacking configuration, respectively, and $A$ is the $xy$ unit cell area. Contour plots of $\Delta E_{(x^*, y^*)}$ for the nine $M_2X$ bilayer models are shown in **Fig. 2b**. These

PESs exhibit periodicity in the 2D ($x^*$, $y^*$) displacement during sliding. Additionally, the PES for bilayer graphene is calculated for comparison (see Supplementary **Fig. 2**). The arrangement of these PESs in **Fig. 2b** reflects a trend in which the electronegativity of the cations and anions decreases horizontally from left to right and vertically from bottom to top, respectively. In tribology, an important metric is the maximum energy barrier during sliding, $\Delta E_{max}$, defined as the difference between the maximum and minimum energy per unit area along the sliding trajectory, i.e., the maximum value of $\Delta E_{(x^*, y^*)}$. As we move towards the lower right corner ($Ba_2N$ bilayer), the PES becomes smoother, with a corrugation amplitude ($\Delta E_{max}$) as low as 0.95 meV/Å$^2$, significantly lower than the value of 3.36 meV/Å$^2$ for the graphene bilayer. Moreover, the calculated $\Delta E_{max}$ shows a notable dependence on the ionic size ratio ($R_{X/M}$) between the anion X and the cation M. $M_2X$ bilayers with smaller $R_{X/M}$ values exhibit smaller $\Delta E_{max}$ (see Supplementary **Fig. 3** and **Table 1**). Notably, when the ionic size ratio drops below 1.3 (e.g., for $Ba_2N$), the calculated $\Delta E_{max}$ is lower than that of the graphene bilayer.

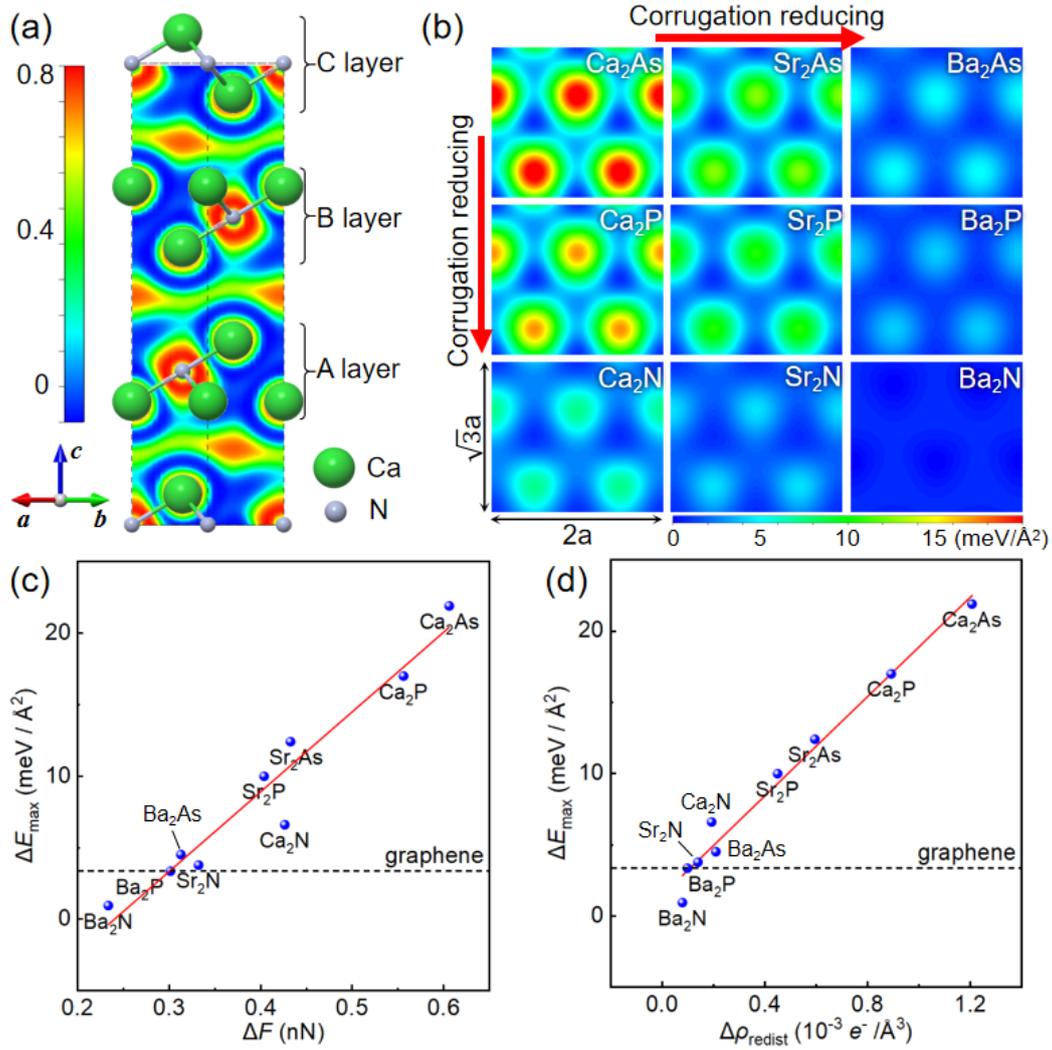

**Fig. 2.** (a) Side view of the bulk unit cell of $Ca_2N$ electride with the calculated electron localization function (ELF). (b) Calculated 2D PESs for the sliding of nine $M_2X$ electrides in bilayer form. (c) Linear correlation between the sliding energy barrier $\Delta E_{max}$ of $M_2X$ bilayers and the Coulombic repulsion difference $\Delta F$ of nearest cations between the minimum- and maximum-energy configurations. (d) Linear correlation between $\Delta E_{max}$ and the interlayer average charge redistribution ($\Delta \rho_{redist}$).

We examine the correlation between $\Delta E_{max}$ and $\Delta F$, where $\Delta F$ represents the difference in Coulombic repulsion between the two cationic layers in their minimum- and maximum-energy configurations. This repulsion is expressed as $F=kQ^2/d^2$, where $k$ is the Coulomb's constant, $d$ denotes the shortest distance between two cations in adjacent layers, and $Q$ is the net charge of the cations, determined through Bader charge analysis (see Supplementary **Table 2**).[29, 30]

In electrides, accurately estimating the cation charge requires incorporating the contribution of interstitial electrons. To achieve this, we perform Bader charge analysis on the pre-converged total charge density, referencing the projected charge density and employing a pseudo-atom model to account for interlayer anionic electrons. Additional computational details are provided in the Supporting Information. Using this approach, we identify the minimum- and maximum-energy configurations across various $M_2X$ bilayers, and the Coulombic repulsion difference $\Delta F$ can be derived from the geometric relationship between adjacent cations:

$$\Delta F = kQ^2 \left( \frac{1}{z_0^2} - \frac{1}{a^2/3 + z_0^2} \right)$$

(2)

We observe a strong linear correlation (Pearson correlation coefficient 0.97) between $\Delta F$ and $\Delta E_{max}$ for the nine $M_2X$ bilayers, as depicted in **Fig. 2c**. This straightforward model provides a convenient method for predicting the frictional properties of anti-$CdCl_2$ 2D electrides based on the net charge of the cations ($Q$) and intrinsic structural factors ($a, z_0$).

Ultralow friction energy barriers in layered materials typically arise from weak interlayer interactions.[31, 32] However, in $Ba_2N$ electride, the adhesion energy between consecutive layers significantly exceeds that of graphite (48.06 compared to 16.96

meV/Å$^2$, see Supplementary **Table 1**), suggesting a crucial role for interlayer electrons in 2D electride sliding. It has been proposed that energy corrugation in friction originates from the non-uniform distribution of charge density in the interface region or from the evolution of charge density in dynamic stacking configurations.[33, 34, 35, 36, 37] Consequently, we conduct a detailed analysis of the relationship between the redistribution of interlayer electrons and the sliding energy barrier. To quantify the accumulation of interlayer electrons, we compute the difference between the total charge density $\rho_{tot}$ of the M$_2$X bilayer and the sum of the charge densities $\rho_{upp}$ and $\rho_{low}$ of the isolated upper and lower layers: $\rho_{diff} = \rho_{tot} - (\rho_{upp} + \rho_{low})$. Contour plots of $\rho_{diff}$ for the Ca$_2$N bilayer are shown in Supplementary **Fig. 4**. Charge density profiles along $z$, calculated by integrating $\rho_{diff}$ over the surface area of the simulation cell $A$ ($\rho_{diff,xy}(z) = \iint_A \rho_{diff}(x,y,z)dxdy$), reveal significant charge transfer from the cationic layers to the interlayer region upon bilayer formation. We evaluate the amount of interlayer charge redistribution $\rho_{redist}$,[31] by integrating $\rho_{diff,xy}(z)$ over the interlayer region ($z_0$, see Supplementary **Fig. 1**) and normalizing by the interlayer volume:

$$\rho_{redist} = \frac{1}{Az_0} \int_{-\frac{z_0}{2}}^{\frac{z_0}{2}} |\rho_{diff,xy}(z)| dz$$

(3)

We investigate the impact of the different stacking configurations on $\rho_{diff,xy}(z)$ by computing the interlayer charge evolution upon sliding $\Delta\rho_{redist}$, i.e., the difference between $\rho_{redist}$ in the minimum- and maximum-energy stacking configurations. **Figure 2d** shows $\Delta E_{max}$ as a function of $\Delta\rho_{redist}$ for the nine M$_2$X electrides, along with a linear fit showing a high Pearson correlation coefficient of 0.99. This indicates a strong correlation between the interlayer charge evolution and energy barriers upon sliding.

Notably, all values of $\Delta\rho_{\text{redist}}$ are positive, indicating an increase in interlayer charge redistribution as the bilayer transitions from the minimum- to the maximum-energy stacking configuration. The higher the increase in the interlayer charge redistribution, the larger the energy barrier generated during sliding, and vice versa. Consequently, $Ba_2N$, with the smallest sliding energy barrier, exhibits the lowest increase in interlayer charge redistribution. The observed correlation between $\Delta\rho_{\text{redist}}$ and $\Delta E_{\text{max}}$ suggests that electron redistribution acts as the primary energy dissipation pathway during sliding. These findings offer valuable insights into leveraging the interlayer electrons of 2D electrides to tune their frictional properties.

**Inducing superlubricity in 2D electrides through incommensurate contact.** Previous research indicates that interlayer friction in rigid crystal interfaces can be influenced by lattice mismatch.[10, 13, 38] Misaligned lattices persist during sliding, leading to reduced energy corrugation. Although lattice mismatch can be quantified using geometric descriptors, other significant effects accompanying lattice twisting, such as in-plane/out-of-plane elastic deformations and thermal corrugation, are often overlooked. Although *ab initio* simulations can capture these phenomena, they are computationally prohibitive for large systems. To address the challenge of simulating moiré superlattices containing over 10,000 atoms, we adopted the DeePMD-kit framework,[39, 40] utilizing interatomic forces generated by neural networks trained on *ab initio* data. This approach achieves DFT-level accuracy and has been validated in large systems containing ~11,000 atoms per moiré unit cell.[41, 42] Our machine learning force field (MLFF) demonstrates high accuracy with force and energy errors below 58 meV/Å and 3 meV per atom, respectively (see Supplementary **Fig. 5**), enabling direct simulation of twisted bilayer $Ba_2N$'s dynamic behavior.

We construct 194 distinct Ba$_2$N bilayer models with twist angles ranging from 0° to 60° using coprime integer pairs (*m*, *n*) to define twisted superlattices through rotational stacking of monolayers (see Supplementary **Fig. 6**). At a twist angle of 60° (antiparallel-stacked, AP), the alignment of the two atomic layers in the bilayer model is opposite to that of the initial structure at 0° (parallel-stacked, P). The stacking configurations during interlayer sliding for both stacking orientations are illustrated in **Fig. 3a**. Each configuration is labeled according to the pair of atoms facing each other across the interface, with the initial configuration denoted as XX'(P).

The interlayer sliding energy profiles for both P (0°) and AP (60°) stacking orientations are calculated using first-principles methods. As shown in **Fig. 3b**, the energy barriers differ significantly: 15 meV for P and only 4 meV for AP. This reduction in the AP case is attributed to the unique alignment of cationic sublayers, where opposing lattice vectors effectively cancel out periodic potential mismatches, leading to a flatter sliding PES. The AP stacking orientation also minimizes the Coulombic repulsion between the cationic layers. Notably, the P stacking orientation features two high-energy configurations (MM'(P) and XM'(P)), while the AP stacking orientation exhibits two nearly degenerate low-energy configurations (MX'(AP)) and (XM'(AP)). These stacking-dependent energy variations are critical for understanding the energetics and structural stability of twisted superlattices.

The relaxed structures of Ba$_2$N bilayers with small twist angles (0° < θ < 2° and 58° < θ < 60°) reveal prominent bright domains characterized by dislocation networks (see **Figs. 3c** and **3d**). In the bilayer with P orientation and a small twist angle (*θ* = 1.08°, left panel of **Fig. 3c**), the lattice reconstructs into a hexagonal array of stacking domains (XX'(P)), which are separated by dislocation networks alternating between high-energy configurations (MM'(P) and XM'(P)). Conversely, in the bilayer with AP orientation and a small twist angle ($\theta = 58.92°$, left panel of **Fig. 3d**), each moiré supercell contains two triangular domains corresponding to low-energy configurations (MX'(AP) and XM'(AP)), while the dislocation networks adopt the high-energy configuration (MM'(AP)). This overall tendency toward low-energy states drives the expansion of bright domains outwards from low-energy centers. Furthermore, at small twist angles, the mismatch between adjacent low-energy domains leads to the accumulation of dislocations in intermediate high-energy regions, forming dislocation walls that separate the domains.

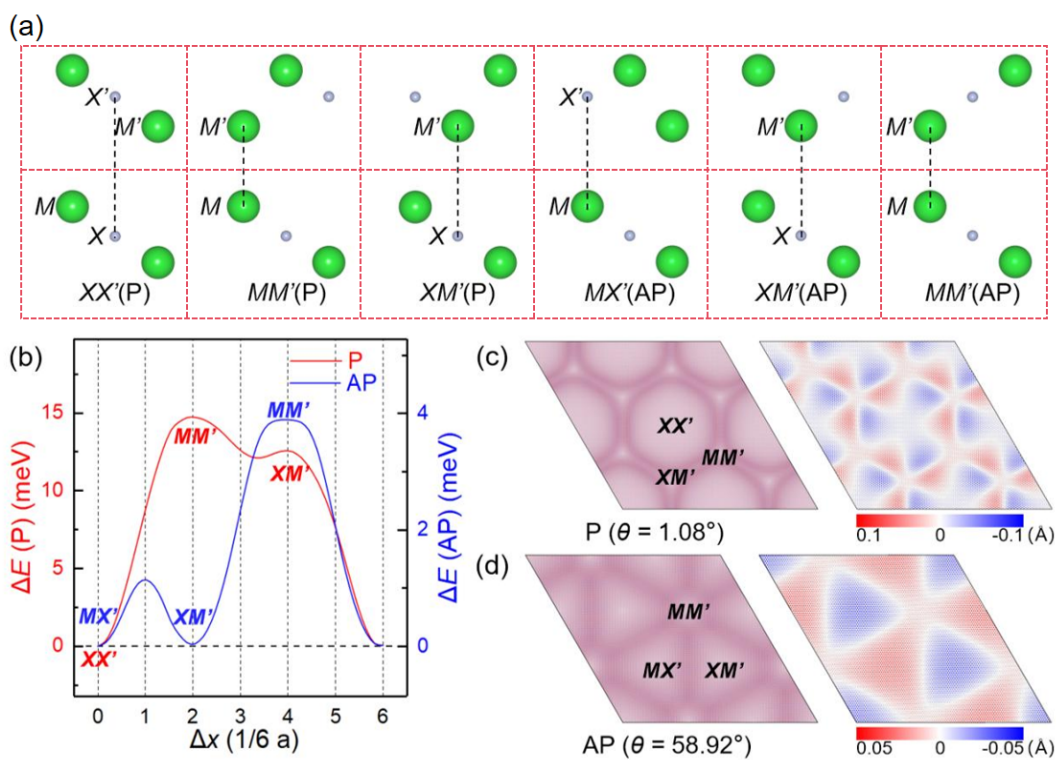

**Fig. 3.** (a) Schematic representation of the typical stacking configurations of bilayer Ba$_2$N with twist angles of 0° and 60°. The green and cyan balls represent Ba and N atoms, respectively. (b) Energy variation of typical stacking configurations during sliding of the bilayer. The sliding direction is along the parallel (P) or antiparallel (AP) orientations of the initial bilayers. (c) Moiré patterns (left) and color maps of the z-

coordinate of the atoms (right) in one cationic layer of relaxed Ba$_2$N superlattices with twist angles of 1.08° and (d) 58.92°. The color scheme for the maps of *z*-coordinate ranges from red (0.1 Å in (c) or 0.05 Å in (d)) to blue (-0.1 Å (c) or -0.05 Å (d)).

Moreover, the enlargement of low-energy areas causes atomic layers to buckle out of plane uniformly across all layers in bilayer structures, regardless of the twist angle. The distribution of these buckles is influenced by perfect domains and dislocation networks. In the P orientation bilayer structure ($\theta$ = 1.08°, right side of **Fig. 3c**), a hexagonal domain is divided into six separate triangular buckles, distributed alternately near the domain walls. Conversely, in the AP orientation bilayer structure ($\theta$=58.92°, right side of **Fig. 3d**), two equally sized triangular domains in low-energy configurations MX'(AP) and XM'(AP) exhibit opposite out-of-plane buckling. No buckling is observed in the atomic layers atop the dislocation network in either scenario. These buckling patterns, which are influenced by low-energy perfect domains and high-energy dislocation networks, significantly affect the frictional properties of the bilayers.

For dynamic friction analysis, we perform molecular dynamic simulations on 36 selected representative models (the selection criteria are detailed in Supplementary **Fig. 7**). The twist angles and lattice constants of these 36 models are listed in Supplementary **Table 3**. **Figure 4a** shows the characteristic lattice structure at $\theta$ = 21.79°. **Figure 4b** shows the simulation setup: spring-based constraints are applied to the basal layer atoms of the substrate. The sliding layer is connected to a virtual atom moving at $v_0$ = 10 m/s along the lattice diagonal (**Fig. 4a**). A uniform normal load is applied to each top Ba atom of the sliding layer. For all systems, constant-temperature molecular-dynamics simulations are performed for 1 ns at 300 K using a Nose-Hoover thermostat. The frictional performance is evaluated through the calculation of the lateral shear strength

($\tau = F_k / A$), where $F_k$ represents the time-averaged kinetic friction force and $A$ denotes the contact area. The calculated shear strengths $\tau$ of the 36 models are listed in Supplementary **Table 3**.

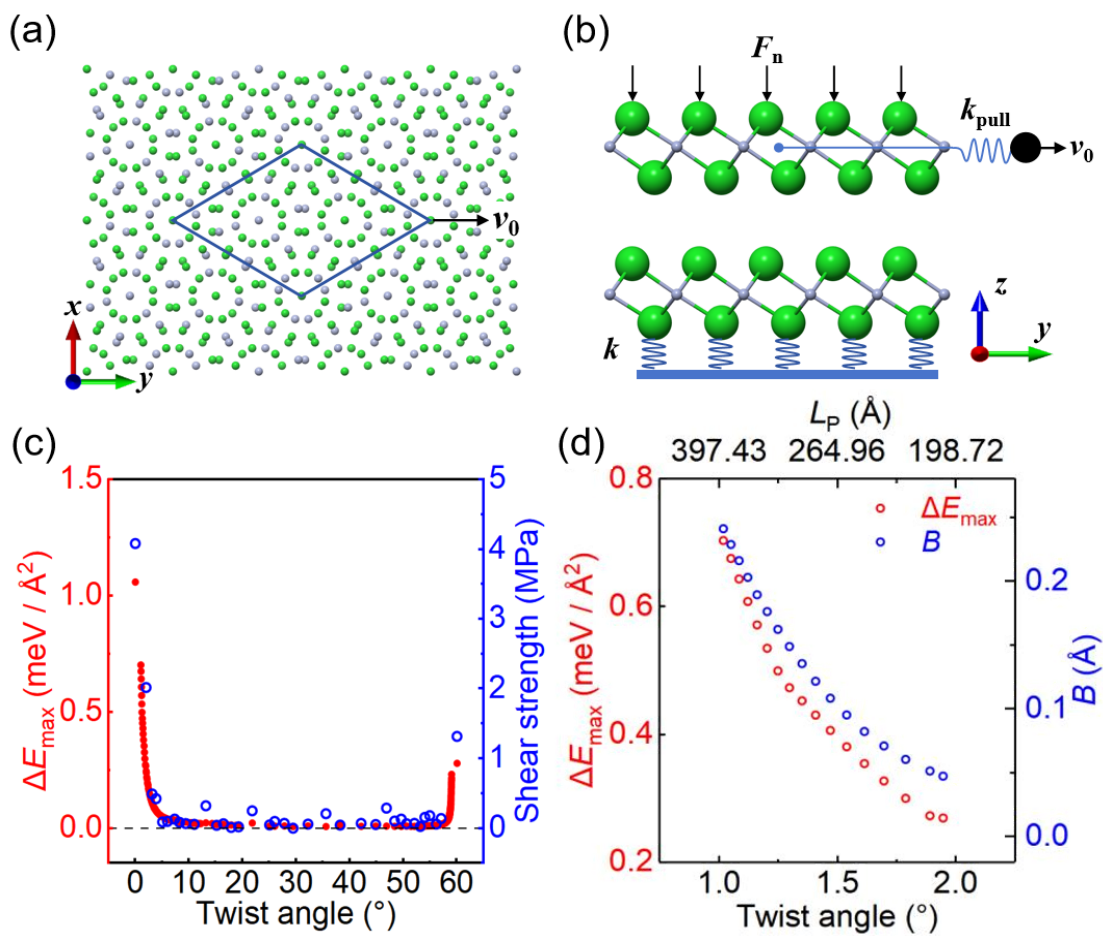

**Fig. 4**. (a) Top view of the bilayer $Ba_2N$ model for a twist angle of $\theta = 21.79°$; (b) Schematic illustration of the simulation setup: Basal Ba atoms in the substrate layer are fixed by springs with elastic constant $k$ to maintain their initial positions. A virtual atom (black sphere) connects to the sliding layer through a spring (elastic constant $k_{pull}$), moving with a constant velocity $v_0$ parallel to the sliding direction. Ba and N atoms are represented by green and blue spheres, respectively. (c) Sliding energy barrier $\Delta E_{max}$ (red dots) and shear strength (blue dots) of bilayer $Ba_2N$ as a function of the twist angle in the range $0° < \theta < 60°$. (d) Sliding energy barrier $\Delta E_{max}$ (red circles) and out-of-plane

buckling $B$ (blue circles) of bilayer Ba$_2$N for small twist angles in the range $0° < \theta < 2°$.

**Figure 4c** presents the dependence of the maximum energy barriers ($\Delta E_{max}$) and lateral shear strengths ($\tau$) on the twist angles under zero normal load conditions. Commensurate lattices at $\theta = 0°$ and $60°$ exhibit energy barriers of 1.06 and 0.28 meV/Å$^2$, respectively. These barriers undergo a dramatic reduction within two narrow range of twist angles ($0° < \theta < 2°$ and $58° < \theta < 60°$), collapsing to ultralow values ($<$ 0.25 meV/Å$^2$) across incommensurate angles ($2° < \theta < 58°$), which strongly suggests the presence of structural superlubricity. This behavior mirrors the observed shear strength reduction approaching $0°$ and $60°$, demonstrating an angular correlation between energy landscape features and kinetic friction. The observed ultralow friction regime is consistent with previous reports on other vdW heterostructures, including twisted graphene,[38] MoS$_2$,[10] and heterogeneous structures,[13] where complete lattice mismatch leads to an ultra-flat PES through destructive interference of atomic corrugations.

Notably, the residual energy barriers at small twist angles ($\theta < 2°$; $\theta > 58°$) originate from structural buckling in Ba$_2$N bilayers. Our analysis identifies two competing mechanisms: (1) out-of-plane buckling amplitudes ($B$), quantified as the maximum z-axis displacement difference within atomic layers, and (2) the in-plane perfect domain size ($L_P$), determined by the lateral distance between two adjacent lowest-energy stacking configurations (see **Figs. 3c** and **3d**). **Figure 4d** depicts the relationships between the twist angle ($\theta$), sliding energy barrier ($\Delta E_{max}$), in-plane perfect domain size ($L_P$), and out-of-plane buckling amplitude ($B$) for small twist angles ($1° < \theta < 2°$). For $\theta$ $< 2°$, both $\Delta E_{max}$ and $B$ display monotonic decreases with increasing $L_P$ and $\theta$. The observed inverse correlation demonstrates that contraction of the in-plane perfect domain imposes geometric constraints on the out-of-plane deformations, suppressing out-of-plane atomic displacements during friction and thereby reducing PES corrugation. When the twist angle exceeds 2°, the out-of-plane buckling amplitudes diminish below 0.05 Å, inducing a rigid incommensurate lattice behavior in bilayer systems. This transition drives the PES to its global minimum, marking the onset of structural superlubricity. For practical applications, stringent angular control ($\theta > 2°$ misalignment) is required to preserve incommensurate interfaces while avoiding spontaneous reconstruction into commensurate domains.

**Inducing superlubricity in 2D electrides through commensurate contact.** Moreover, the exceptional frictional dynamics of pristine $Ba_2N$ bilayers ($\theta = 0°$) attracted our interest, exhibiting an ultralow shear strength of 4 MPa (**Fig. 4c**), which is substantially lower than the intrinsic shear strength of AB-stacked graphite of 62 MPa.[43] These properties endow $Ba_2N$ with significant potential for applications in nanomechanical systems and superlubricity-enabled materials. The ultralow friction behavior originates from the synergistic effect of a zigzag-type sliding pathway. As illustrated in **Fig. 5a**, the periodic evolution of the centroid of the sliding layer along the *x*-direction (normalized to the lattice constant *a*) exhibits quantized transitions in increments of 0.5*a*, indicating movement along energy-minimized paths. By consistently following these minimum-energy trajectories, the system effectively avoids large-scale energy dissipation, thereby enabling ultralow friction during sliding.

Given the prevalence of high-load conditions in practical tribological interfaces, we further investigated the frictional behavior of $Ba_2N$ under applied normal loads. To monitor the energy variations at the origin of friction, we performed first-principles calculations to track the load-dependent energy difference ($\Delta E$) between AB and AA stacking configurations, corresponding to the maximum and minimum of the sliding PES, respectively. **Figures 5b** and **5c** show the tribological behavior of commensurate Ba$_2$N bilayers ($\theta = 0°$) under varying normal loads, revealing two distinct friction regimes.

Regime I (0 - 2.3 GPa): The shear strength demonstrates a counterintuitive reduction with increasing normal load (**Fig. 5b**), while DFT calculations reveal that the energy barrier $\Delta E$ initially increases weakly (**Fig. 5c**). This contradiction can be attributed to the evolution of the PES. Under load-free conditions, the bilayer tends to bypass AB stacking and slide along the zigzag-type minimum energy path (**Fig. 5a**). The system actually have to overcome the transition-state energy barrier between two AA stackings. With increasing normal load, the PES undergoes dynamic evolution (manifested as a reduction in saddle-point energy), consequently resulting in a reduction of the shear strength. At the critical load of 2.1 GPa, $\Delta E$ vanishes completely, marking the system's transition to a metastable state with barrier-free sliding potential. Meanwhile, the ratio between the shear strength and the normal load reaches 0.001, which is lower than 0.01, which is usually referred to as the superlubricity threshold.[44] Regime II (> 2.3 GPa): Beyond the critical load, $\Delta E$ resurges with inverted polarity (**Fig. 5c**), indicating a fundamental inversion of the stacking stability hierarchy - the formerly high-energy AB configuration becomes energetically favorable. This inversion drives the restored positive correlation between the shear strength and load (**Fig. 5b**), accompanied by a linear growth of $|\Delta E|$.

First-principles charge density difference analysis (**Fig. 5d**) unveils topological reconstruction of interlayer electron clouds under high loads. At ambient pressure (0 GPa), both AA and AB stacking configurations exhibit moderate charge redistribution.

The AA stacking configuration exhibits significantly enhanced interlayer electron accumulation ($\rho_{redist}$ = 0.333 e$^-$/Å$^3$) compared to AB stacking ($\rho_{redist}$ = 0.329 e$^-$/Å$^3$), as evidenced by the spatially extended red isosurfaces in the interlayer region. This electronic enrichment enhances interlayer Coulomb repulsion through charge compensation effects, resulting in an energetically favorable state with 0.95 meV/Å$^2$ lower total energy than AB stacking. Under 2.3 GPa compression, dramatic electronic reconstruction emerges: For AA stacking, intensified charge accumulation develops at interfacial coordination zone, forming an anisotropic electron sphere that spans interlayer gap. For AB stacking, pressure induces asymmetric charge polarization, with alternating accumulation/depletion domains aligning along the shear direction. This distinct behavior can be attributed to the different nitrogen alignments: in AA stacking, the vertically aligned N atoms create symmetric interfacial electric fields that reinforce directional charge confinement, whereas in AB stacking, the staggered N positions induce competing dipole interactions that drive directional charge redistribution. Notably, interfacial charge forms novel charge-compensation channels (red filaments between layers), geometrically matching the sliding pathway observed in barrier-free sliding regime, which effectively screens the interlayer Coulomb repulsion. The difference in the charge density quantitatively demonstrates a pressure-induced inversion of the charge transfer hierarchy, with the AA stacking configuration exhibiting lower interlayer electron accumulation ($\rho_{redist}$ = 0.353 e$^-$/Å$^3$) compared to the AB stacking ($\rho_{redist}$ = 0.397 e$^-$/Å$^3$). This electronic inversion correlates directly with the metastable stacking state transition observed in **Fig. 5c**, confirming that pressure-tuned charge redistribution governs the interlayer interaction potential landscape. This frictional collapse behavior is also observed in rare gas/metals and MoS$_2$ vdW friction systems.[6, 45]

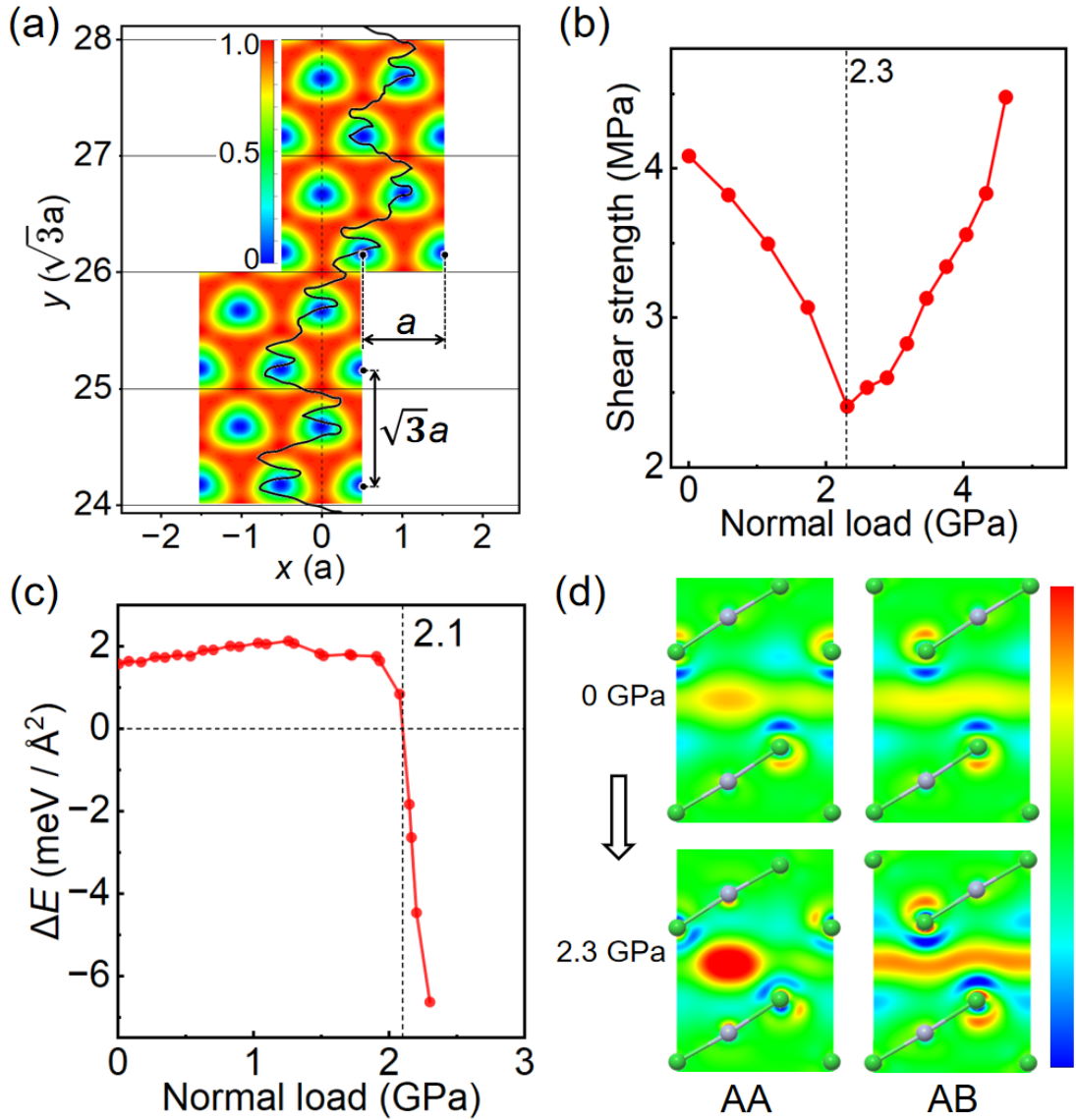

**Fig. 5.** (a) Evolution of the centroid's *x* and *y* coordinates of sliding layer (normalized by the lattice parameters *a* and $\sqrt{3}a$, respectively) under load-free conditions (initial position set to (0,0)). The curve shows a zigzag-type sliding path mapping on the PES (meV/Å²). (b) Lateral shear strength as a function of the normal load. (c) The Energy difference ($\Delta E$) between AB and AA stacking configurations as a function of the normal load. (d) The Difference charge density map on the (110) crystallographic plane of $Ba_2N$ bilayer for AA and AB stacking configurations under 0 and 2.3 GPa. The isosurfaces range from -0.002 (blue, charge depletion) to 0.002 (red, charge accumulation) (units:

e⁻/Å³). The green and gray spheres represent Ba and N atomic positions, respectively.

Unlike incommensurate systems (2° < $\theta$ < 58°) where friction reduction originates from disordered moiré patterns, the $Ba_2N$ commensurate bilayer maintains long-range crystallographic order while achieving complete energy corrugation elimination. The robustness of this state is further enhanced by the strong interlayer adhesion of electrides, which prevents exfoliation under shear stress, a common failure mode in weakly bonded vdW materials. By preserving crystallographic alignment, this mechanism enables precise control over interfacial properties without sacrificing mechanical stability. The observed metastable barrier elimination at critical loading demonstrates a unique pathway for achieving load-adaptive superlubricity in electrides, distinct from conventional structural lubricity mechanisms. These findings establish electrides as a promising platform for designing pressure-responsive smart tribological systems with ultralow friction under extreme loading conditions.

**Tuning the friction of 2D electrides through electron doping. Figures 2d** and **5d** suggest that reducing the interlayer charge redistribution can mitigate the energy barrier associated with the stacking configuration displacement. To explore this phenomenon further, we examine the impact of electron doping on the frictional properties of 2D electrides by varying the electron concentration in the bilayer $Ba_2N$ model. The sliding energy barrier ($\Delta E_{max}$) of bilayer $Ba_2N$, calculated for electron doping levels ranging from 0 to 0.5 e/f.u., is presented in **Fig. 6a**. It can be observed that, as $n_e$ increases, $\Delta E_{max}$ initially decreases significantly, indicating a pronounced sensitivity to electron doping. However, $\Delta E_{max}$ reaches a minimum of 0.58 meV/Å² at $n_e \approx 0.45$ e/f.u., after which it begins to increase as $n_e$ continues to increase.

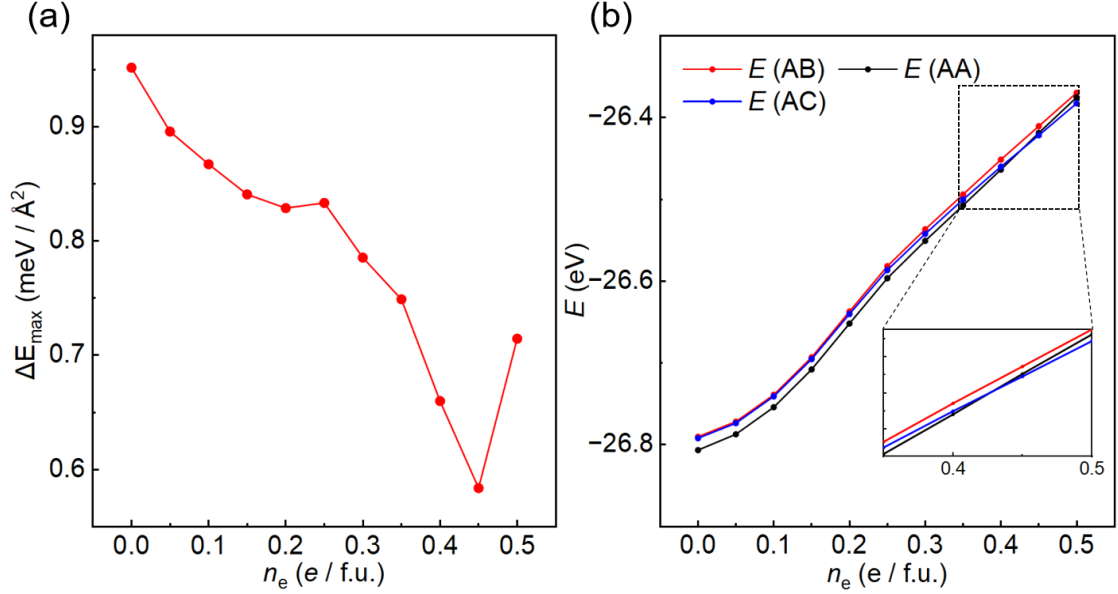

**Fig. 6.** (a) Variation of the sliding energy barrier ($\Delta E_{max}$) for bilayer $Ba_2N$ as a function of the number of doping electrons ($n_e$). (b) Energies of the stacking configurations AB, AA, and AC for bilayer $Ba_2N$ as a function of $n_e$. The inset in (b) highlights that the energy of the AC stacking configuration becomes lower than that of the most stable AA configuration at $n_e \approx 0.45$ e/f.u.

We further investigated the impact of electron doping on the stacking configurations of bilayer $Ba_2N$. **Figure 6b** shows the energies of different stacking configurations—AB, AA, and AC—as a function of the doping electron concentration ($n_e$), where $E(AB)$, $E(AA)$, and $E(AC)$ correspond to the energies of the AB, AA, and AC stacking configurations, respectively. The initial sliding energy barrier is defined by the energy difference between AA and AB. As $n_e$ increases from 0 to 0.5 e/f.u., the stacking energies of all three configurations increase monotonically. Notably, the energy difference between AA and AB decreases, while the energy difference between AC and AB increases. Interestingly, when $n_e$ reaches approximately 0.43 e/f.u., the energy curves of AA and AC intersect, causing the sliding energy barrier to shift to the energy difference between AC and AB, which continues to rise with further electron doping. The intersection in **Fig. 6b** implies an exchange in the energy ordering between stacking configurations, analogous to frictional collapse behaviors observed in gas/metal and MoS$_2$ vdW systems.[6, 45] Notably, the influence of electron doping on the frictional properties of Ba$_2$N can also be observed using a bulk model (Supplementary **Figs. 8** and **9**), providing further evidence for the modulation of friction in 2D electrides. Our findings offer a deeper understanding of the factors that influence stacking configurations in 2D electrides, suggesting that fine-tuning the electron doping concentration could enable control over the preferred stacking motifs and, consequently, the frictional properties of materials. To introduce additional electrons into the Ba$_2$N lattice, one could either apply an external electrostatic field to modulate the Fermi level,[46] or substitute heteroatoms for nitrogen atoms in the Ba$_2$N lattice.[47] The doping-induced friction reduction suggests potential applications in adaptive lubrication systems where friction coefficients can be dynamically tuned through electrostatic gating,[46] creating smart interfaces responsive to operational conditions.

**Discussion**

In summary, we have investigated the tribological properties of 2D electrides using *ab initio* and DPMD calculations. We uncovered the mechanism of atomic-scale friction within these 2D electride systems, considering both structural and electronic factors. The sliding energy barriers of 2D electrides correlate with the Coulombic repulsion difference between the maximum and minimum energy stacking configurations. Contrary to the prevailing belief that systems with high interlayer binding energy exhibit poor lubrication properties, the sliding energy barrier for Ba$_2$N is about three times smaller than that of graphite though the adhesion energy is significantly higher.

DPMD simulations reveal that incommensurate contact configurations in $Ba_2N$ bilayers demonstrate significantly reduced shear strength (< 0.5 MPa). Remarkably, commensurate contact systems under an extra critical pressure (2.3 GPa) exhibit superlubricity, with the ratio between the shear strength and the normal load reaching 0.001. Finally, by leveraging the pivotal role of interlayer electrons, friction in 2D electrides can be significantly reduced through electron doping. These findings highlight the interplay between the tribological, structural, and electronic properties of 2D electrides. They highlight incommensurate contact, commensurate contact, and electron doping as effective strategies for reducing friction in these materials, offering significant potential for cost-efficient fabrication and practical engineering applications.

**Method**

*Ab initio* **calculations.** The *ab initio* calculations were performed based on density functional theory (DFT)[48] using the projector-augmented wave method as implemented in the Vienna Ab-initio Simulation Package (VASP).[49, 50] The exchange-correlation energy was modeled using the generalized gradient approximation (GGA) of Perdew, Burke, and Ernzerhof (PBE).[51] Long-range vdW interactions were added using Grimme's DFT-D3 approach.[52] The wavefunctions were expanded on a plane-wave basis set with a cutoff energy of 520 eV. The Brillouin zone was sampled using an 11×11×11 k-point mesh. Before calculating the static potential energy surface (PES), all atoms were completely relaxed until the forces were smaller than 10 meV/Å. The initial data set was generated from AIMD simulations, which were performed based on canonical (NVT) ensembles with Nosé–Hoover thermostats.[53, 54]

**Bader charge analysis.** The Bader method considers the zero-flux surface of the charge density to divide the molecular space into atomic volumes known as Bader basins.[29, 30] In electrides, the volume of the Bader basin around the interlayer anionic electrons where the electron density is maximum, known as a non-nuclear attractor, is estimated to be an empty sphere (pseudo atom). The central coordinates of the interlayer anionic electrons are all (0, 0, 0.5), which are determined semi-empirically in the unit cell by examining the electron localization function (ELF) regions separated from the atoms (see Supplementary **Fig. 1**). Here, the Bader charge analysis was carried out using the pre-converged total charge density with reference to the projected charge density, which employs a substitution of pseudo atom for interlayer anionic electrons.

**MLFF Model Training.** DeePMD-kit software[39, 40] was used to fit the neural network for atomic interactions within a deep learning framework. The embedding network employed had three hidden layers with 25, 50, and 100 nodes, respectively, while the fitting network used three hidden layers, each with 240 nodes. A cutoff radius of 7.0 Å was applied. The learning rate decayed exponentially from 0.001 every 2000 steps, eventually reaching $3.51 \times 10^{-8}$ after 1,000,000 training steps, with start and limit scaling factors of energy and force set at 0.02, 1, 1000, and 1, respectively.

**Deep Potential Molecular Dynamics.** All DPMD simulations based on MLFF were performed using Large-scale Atomic/Molecular Massively Parallel Simulator (LAMMPS)[55] in the NVT ensemble. Periodic boundary conditions were applied in the $x$ and $y$ directions, whereas non-periodic and fixed boundary conditions were applied in the $z$ direction. The time step was set to 1 fs. The conjugate gradient (CG) energy minimization method with a force criterion of $10^{-12}$ eV/Å was used to pre-equilibrate the initial models. Trajectory analyses were performed with OVITO software.[56]

**Supporting Information**

The Supporting Information is available free of charge at [url].

Model and sliding PES for the 2D electrides bilayer; Coulombic repulsion difference; interlayer charge redistribution; construction of the machine learning force field ; construction of the twisted superlattice models ; DPMD simulation protocol and the shear strength; and simulation of electron doping in the bulk model .

**Acknowledgments**

This work was supported by the National Key Research and Development Program of China (2023YFB3003005). MCR gratefully acknowledges the ERC-SLIDE project (Grant agreement No. 865633).

**Contribution**

J.W. and J.Q. proposed the concept and designed the DFT calculations. G.R and M.C.R designed the MD simulations. J.Q. conducted the calculations and simulations, and drafted the original manuscript. J.W., G.M, G.R and M.C.R reviewed and edited the manuscript. J.W. supervised the project. All the authors contributed to the overall scientific interpretation and edited the manuscript.